# Impact of Rotation on the Retarded Interaction between a Permanent Dipole Particle and a Polarizable Surface

A.A. Kyasov and G. V. Dedkov

Nanoscale Physics Group, Kabardino-Balkarian State University, Nalchik, 360004, Russia

We have calculated components of torque and the interaction energy of small rotating particle with a permanent dipole moment in the case where the rotation axis is perpendicular to the surface and the dipole axis is inclined to it. The retardation effects were taken into account. An important property of this system is its quasistationarity. When the dipole axis coincides with the rotation axis, the particle does not experience damping and may revolve for infinitely long time. When the axis of dipole particle is inclined to the surface, the situation is possible where the particle is repelled off the surface and its spin increases.

## 1. Introduction

Light interaction with matter is known to produce mechanical rotation of even micron-sized neutral particles [1,2]. The reverse situation where the radiation can be emitted by rotating particles was considered in [3]. The impact of the uniform motion of particles on the van der Waals and Casimir forces was studied by us in [4], while in [5] we have studied the effect of rotation on these interactions. Quite recently, the frictional torque on a particle rotating near a surface has been calculated by us and Zhao et. al. [5,6]. In particular, it was shown that the stopping time of rolling particle in the near field of the surface is by 5 to 9 orders of magnitude lower than for rolling in vacuum and at uniform motion near the surface.

As for the dipole particles with a permanent dipole moment, rotational effects of non-thermal origin can also change their interactions with micro- and macroscopic bodies. The purpose of this paper is to study the influence of rotation on the interaction of permanent dipoles with the surface of dielectric/metallic slab with the dielectric permittivity $\varepsilon(\omega)$. Even in this rather simple situation we obtain new effects if the rotation frequency is high enough for retardation effects to appear. In particular, the surface of dielectric/metal may attract or repel the rotating particle, while the torque may exert frictional and accelerating action. These features should be taken into account when studying particle trapping close to the surface in vacuum [7,8], in dynamics of coagulating dust (cluster) particles and in designing NEMS.

## 2. Theoretical consideration

Consider the geometrical configuration shown in Fig. 1. A point –dipole particle with a permanent electrical dipole moment **d**′ in the own coordinate system $\Sigma'$ of the dipole rotates with the angular velocity $\Omega$ around the $Z, Z'$ axis. The second coordinate system $\Sigma$ is related to the resting surface of a slab that is characterized by the frequency-dependent dielectric permittivity $\varepsilon(\omega)$. In the case shown in Fig. 1, the components of dipole moment **d** in $\Sigma$ are given by

$$
\begin{aligned}
d_x &= d_x' \cos\Omega t - d_y' \sin\Omega t \\
d_y &= d_x' \sin\Omega t + d_y' \cos\Omega t \\
d_z &= d_z'
\end{aligned}
\quad (1)
$$

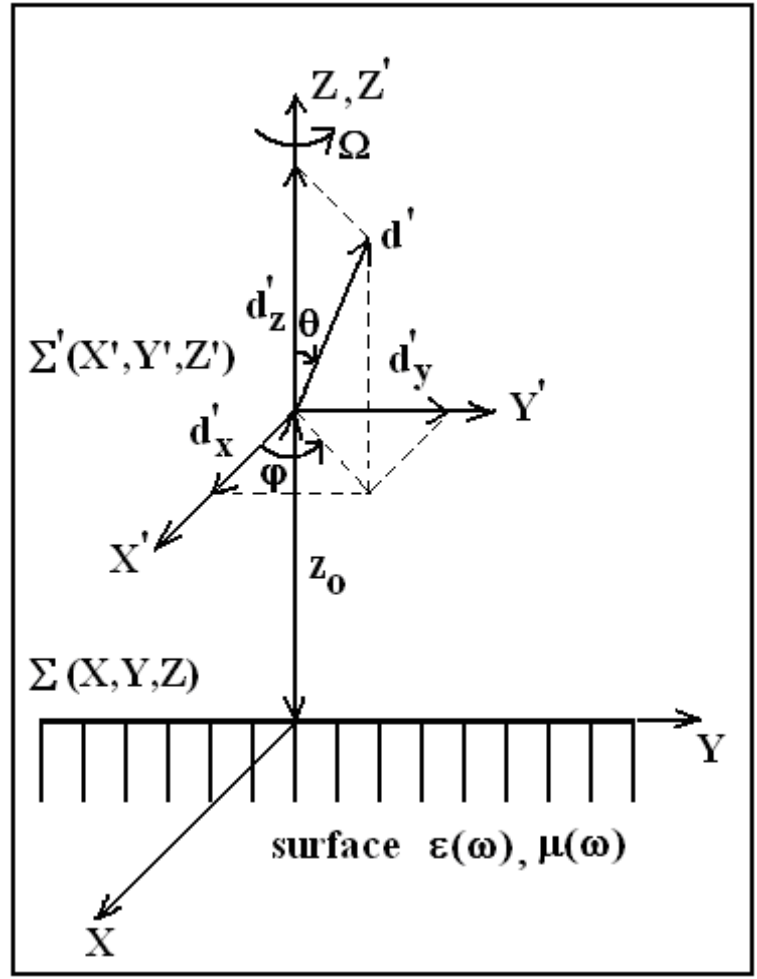

Fig. 1 Geometrical configuration and coordinate systems used.

Using the Fourier representation, direct and inverse Fourier transforms of **d**($t$) are

$$\mathbf{d}(t) = \int_{-\infty}^{+\infty} \frac{d\omega}{2\pi} \mathbf{d}(\omega)\exp(-i\omega t),$$
$$\mathbf{d}(\omega) = \int_{-\infty}^{+\infty} dt\, \mathbf{d}(t)\exp(i\omega t) \tag{2}$$

With allowance for (1), (2), the components of **d** in reference frame $\Sigma$ take the form

$$d_x(\omega) = \pi\left[\delta(\omega^+)(d_x' + id_y') + \delta(\omega^-)(d_x' - id_y')\right],$$
$$d_y(\omega) = \pi\left[-i\delta(\omega^+)(d_x' + id_y') + i\delta(\omega^-)(d_x' - id_y')\right], \tag{3}$$
$$d_z(\omega) = 2\pi\delta(\omega)d_z'$$

where $\omega_\pm = \omega \pm \Omega$.

The starting equations for the particle-surface interaction energy and the torque (in the reference frame $\Sigma$) are given by

$$U(z_0) = -\frac{1}{2}\mathbf{d}(t)\mathbf{E}^{in}(\mathbf{r}_0,t) \tag{4}$$

$$\mathbf{M} = \mathbf{d}(t) \times \mathbf{E}^{in}(\mathbf{r}_0,t) \tag{5}$$

where $\mathbf{r}_0 = (0,0,z_0)$ and $\mathbf{E}^{in}(\mathbf{r}_0,t)$ denotes the induced electric field of the surface. The dependence on time in the right-hand sides of (4), (5) is associated with the rotation of the particle. It is convenient to write $\mathbf{E}^{in}(\mathbf{r}_0,t)$ as the Fourier expansion over the time and the two-dimensional wave vector **k** (parallel to the surface)

$$\mathbf{E}^{in}(\mathbf{r},t) = -\int \frac{d\omega}{2\pi} \frac{d^2k}{(2\pi)^2} \mathbf{E}^{in}_{\omega\mathbf{k}}(z)\exp\bigl(i(k_x x + k_y y - \omega t)\bigr) \tag{6}$$

Substituting (2) and (6) into (4),(5) at the location point of the dipole, $\mathbf{r}_0 = (0,0,z_0)$, yields

$$U(z_0) = -\frac{1}{2}\int \frac{d\omega\, d\omega'}{2\pi\, 2\pi} \frac{d^2k}{(2\pi)^2} \mathbf{d}(\omega')\mathbf{E}^{in}_{\omega\mathbf{k}}(z_0)\exp\bigl(-i(\omega+\omega')t\bigr) \tag{7}$$

$$\mathbf{M} = \int \frac{d\omega\, d\omega'}{2\pi\, 2\pi} \frac{d^2k}{(2\pi)^2} \mathbf{d}(\omega') \times \mathbf{E}^{in}_{\omega\mathbf{k}}(z_0)\exp\bigl(-i(\omega+\omega')t\bigr) \tag{8}$$

The components of $\mathbf{E}^{in}{}_{\omega\mathbf{k}}(z_0)$ are determined from the solution to the Maxwell equations with a point-like dipole source, provided that the corresponding boundary conditions are satisfied at $z = 0$ (see, for instance, [9]):

$$E_{x,\omega\mathbf{k}}(z_0) = \frac{2\pi}{q_0}\exp(-2q_0 z_0) \cdot$$
$$\cdot\left[ik_x\bigl(ik_x d_x(\omega) + ik_y d_y(\omega)\bigr)\bigl(\Delta_m(\omega) - \Delta_{em}(\omega)\bigr) - ik_x q_0 \Delta_e(\omega) d_z(\omega) + \frac{\omega^2}{c^2}\Delta_m(\omega) d_x(\omega)\right]$$

$$E_{y,\omega\mathbf{k}}(z_0) = \frac{2\pi}{q_0}\exp(-2q_0 z_0) \cdot$$
$$\cdot\left[ik_y\bigl(ik_x d_x(\omega) + ik_y d_y(\omega)\bigr)\bigl(\Delta_m(\omega) - \Delta_{em}(\omega)\bigr) - ik_y q_0 \Delta_e(\omega) d_z(\omega) + \frac{\omega^2}{c^2}\Delta_m(\omega) d_y(\omega)\right] \quad (9)$$

$$E_{z,\omega\mathbf{k}}(z_0) = \frac{2\pi}{q_0}\exp(-2q_0 z_0) \cdot$$
$$\cdot\left[\bigl(ik_x d_x(\omega) + ik_y d_y(\omega)\bigr)\frac{\bigl(k^2 \Delta_{em}(\omega) - q_0^2 \Delta_m(\omega)\bigr)}{q_0} + k^2 \Delta_e(\omega) d_z(\omega)\right]$$

$$q_0 = \sqrt{k^2 - \omega^2/c^2},\ q = \sqrt{k^2 - \varepsilon(\omega)\mu(\omega)\omega^2/c^2},\ \Delta_e(\omega) = \frac{q_0\varepsilon(\omega) - q}{q_0\varepsilon(\omega) + q},$$
$$\Delta_m(\omega) = \frac{q_0\mu(\omega) - q}{q_0\mu(\omega) + q},\ \Delta_{em}(\omega) = \frac{2q_0^2\bigl(\varepsilon(\omega)\mu(\omega) - 1\bigr)}{\bigl(q_0\varepsilon(\omega) + q\bigr)\bigl(q_0\mu(\omega) + q\bigr)} \quad (10)$$

The functions $\Delta_e(\omega), \Delta_m(\omega), \Delta_{em}(\omega)$ are related by

$$\Delta_{em}(\omega) = \left(1 - \frac{\omega^2}{k^2 c^2}\right)\bigl(\Delta_e(\omega) + \Delta_m(\omega)\bigr) \quad (11)$$

Substituting (9)—(11) in (7), (8) with allowance for (11) yields

$$U(z_0) = -\frac{d^2 \cos^2\theta}{8 z_0^3}\left(\frac{\varepsilon(0) - 1}{\varepsilon(0) + 1}\right) - \frac{d^2 \sin^2\theta}{4}\mathrm{Re}\left\{\int_0^\infty dk\,\frac{k}{q_0}\exp(-2q_0(\Omega)z_0)\,R(\Omega,k)\right\} \quad (12)$$

$$M_x = \frac{d^2 \sin\theta\cos\theta}{4 z_0^3}\left(\frac{\varepsilon(0) - 1}{\varepsilon(0) + 1}\right)\sin(\Omega t + \varphi) +$$
$$+ \frac{d^2 \sin\theta\cos\theta}{2}\mathrm{Im}\left\{\int_0^\infty dk\,\frac{k}{q_0(\Omega)}\exp(-2q_0(\Omega)z_0)\exp(-i(\Omega t + \varphi))R(\Omega,k)\right\} \quad (13)$$

$$M_y = -\frac{d^2 \sin\theta\cos\theta}{4 z_0^3}\left(\frac{\varepsilon(0) - 1}{\varepsilon(0) + 1}\right)\cos(\Omega t + \varphi) +$$
$$+ \frac{d^2 \sin\theta\cos\theta}{2}\mathrm{Re}\left\{\int_0^\infty dk\,\frac{k}{q_0(\Omega)}\exp(-2q_0(\Omega)z_0)\exp(-i(\Omega t + \varphi))R(\Omega,k)\right\} \quad (14)$$

$$M_z = -\frac{d^2 \sin^2\theta}{2} \mathrm{Im}\left\{\int_0^\infty dk \frac{k}{q_0(\Omega)} \exp(-2q_0(\Omega)z_0) R(\Omega,k)\right\} \qquad (15)$$

$$R(\Omega,k) = \left(k^2 - \frac{\Omega^2}{c^2}\right)\Delta_e(\Omega) + \frac{\Omega^2}{c^2}\Delta_m(\Omega) \qquad (16)$$

Note that $d$ in (12)—(15) denotes, in essence, the dipole moment in the reference frame $\Sigma'$ of the particle, but owing to the condition $\Omega \cdot a \ll c$ this dipole moment is the same in the system $\Sigma$, as well.

Completely analogous calculations for the rotating magnetic moment **m** lead to the same equations (12)—(15) with simple replacements $d \to m, \Delta_e \to \Delta_m, \Delta_m \to \Delta_e, \varepsilon(0) \to \mu(0)$.

So, equations (12)—(15) describe the retarded electromagnetic interactions of permanent electric (magnetic) dipoles with the spin axis being perpendicular to the surface of a slab.

The nonrelativistic limits ($c \to \infty$) of Eqs. (12)--(16) were obtained in [10] and can be easily recovered from the above equations upon setting $c \to \infty$:

$$U(z_0) = -\frac{d^2}{16 z_0^3}\left[\Delta'(\Omega)\sin^2\theta + 2\Delta'(0)\cos^2\theta\right] \qquad (12a)$$

$$M_x = \frac{d^2 \sin\theta\cos\theta}{4 z_0^3} \cdot \left[(\Delta'(0) - \Delta'(\Omega)/2)\sin(\varphi+\Omega t) + 0.5\Delta''(\Omega)\cos(\varphi+\Omega t)\right] \qquad (13a)$$

$$M_y = \frac{d^2 \sin\theta\cos\theta}{4 z_0^3}\left[-(\Delta'(0) - \Delta'(\Omega)/2)\cos(\varphi+\Omega t) + 0.5\Delta''(\Omega)\sin(\varphi+\Omega t)\right] \qquad (14a)$$

$$M_z = -\frac{d^2 \sin^2\theta}{8 z_0^3}\Delta''(\Omega) \qquad (15a)$$

where $\Delta(\Omega) = \frac{\varepsilon(\Omega)-1}{\varepsilon(\Omega)+1}$, and $\Delta'(\Omega), \Delta''(\Omega)$ are the corresponding real and imaginary components.

From (15a) it follows

$$M_\perp = \sqrt{M_x^2 + M_y^2} == \frac{d^2 \sin\theta\cos\theta}{4 z_0^3}\sqrt{[(\Delta'(0) - \Delta'(\Omega)/2)^2 + \Delta''(\Omega)^2/4]} \qquad (17)$$

Thus, only component $M_z$ of the torque describes the damping effect, while the perpendicular component $\mathbf{M}_\perp = (M_x, M_y)$ synchronously rotates with the particle and has such a direction that the dipole axis tends to turn perpendicularly to the surface. The value of $M_\perp$ depends only on the angle $\theta$ and does nod depend on the angle $\varphi$ and time. For the nonrotating dipole, $\Omega = 0$,

Eq. (17) is reduced to $M_\perp = \dfrac{d^2 \sin\theta \cos\theta}{8z_0^3} \Delta'(0)$, and we obtain the usual effect of dipole orientation near the surface of dielectric. Exactly the same properties stem from (13)—(15), though these equations have a more complicated form. The other important features are as follows: i) potential energy of the system and component $M_z$ of the torque explicitly do not depend on time; ii) at $\theta = 0$, frictional and orientational components of the torque disappear and the particle preserves its rotation for infinitely long time. The interaction potential in this case takes the static form

$$U(z_0) = -\frac{d^2}{8z_0^3}\left(\frac{\varepsilon(0)-1}{\varepsilon(0)+1}\right) \tag{18}$$

Another interesting situation occurs in the case $\theta = \pi/2$, where $M_\perp = 0$, but the interaction potential $U(z_0)$ and frictional torque $M_z$ significantly depend on the angular velocity $\Omega$ and the dielectric properties of the surface. It is worth noting that the dipole approximation implies that $a \ll \min(z_0, c/\Omega)$, but, generally speaking, this condition does not noticeably restrict the value of the retardation parameter $\xi = \dfrac{\Omega z_0}{c} = \dfrac{\Omega a}{c}\left(\dfrac{z_0}{a}\right)$. Therefore, general relativistic solution to this problem has real practical significance. According to (12)—(16), the effect of retardation becomes important only at $\theta \neq 0$, otherwise it is absent. Below these features are demonstrated in the case of an ideally conducting surface, the surface of Drude metal, and that of ionic dielectric.

### 3. Numerical examples

*An ideally conducting surface*

This case corresponds to the limit $\varepsilon(\omega) \to \infty$, $\Delta_e(\Omega) = 1, \Delta_m(\Omega) = -1$, while the integrals in (12)—(16) can be calculated explicitly. So, denoting $\lambda = 2\xi = 2\Omega \cdot z_0/c$, we obtain

$$I = \int_0^\infty dk\, k\, \frac{\exp\left(-2\sqrt{k^2 - \Omega^2/c^2}\, z_0\right)}{\sqrt{k^2 - \Omega^2/c^2}} \left[(k^2 - \Omega^2/c^2)\Delta_e(\Omega) + \Omega^2 \Delta_m(\Omega)/c^2\right] =$$
$$= \frac{1}{4z_0^3}\left[(1-\lambda^2)\cos\lambda + \lambda \sin\lambda + i\cdot(\lambda\cos\lambda + (\lambda^2-1)\sin\lambda)\right] \tag{19}$$

Using (12)—(16) and (19) yields

$$M_\perp = \frac{d^2 \sin\theta\cos\theta}{8z_0^3}\left(5 - \lambda^2 + \lambda^4 + 4(\lambda^2 - 1)\cos\lambda - 4\lambda\sin\lambda\right)^{1/2} \tag{20}$$

$$M_z = -\frac{d^2 \sin^2\theta}{8z_0^3}\left(\lambda\cos\lambda + (\lambda^2 - 1)\sin\lambda\right) \tag{21}$$

$$U(z_0) = -\frac{d^2 \cos^2\theta}{8z_0^3} - \frac{d^2 \sin^2\theta}{16z_0^3}\left((1 - \lambda^2)\cos\lambda + \lambda\sin\lambda\right) \tag{22}$$

Equation (21) and Eq. (22) demonstrate striking possibilities of acceleration of rotating dipolar particle and repulsion from the surface at $\theta = \pi/2$. This is clearly seen from Fig. 2a,b showing the normalized dependences of $M_z$ and $U(z_0)$ on $z_0$ in units $(d^2\Omega^3/c^3)$ and $(d^2\Omega^3/2c^3)$, respectively.

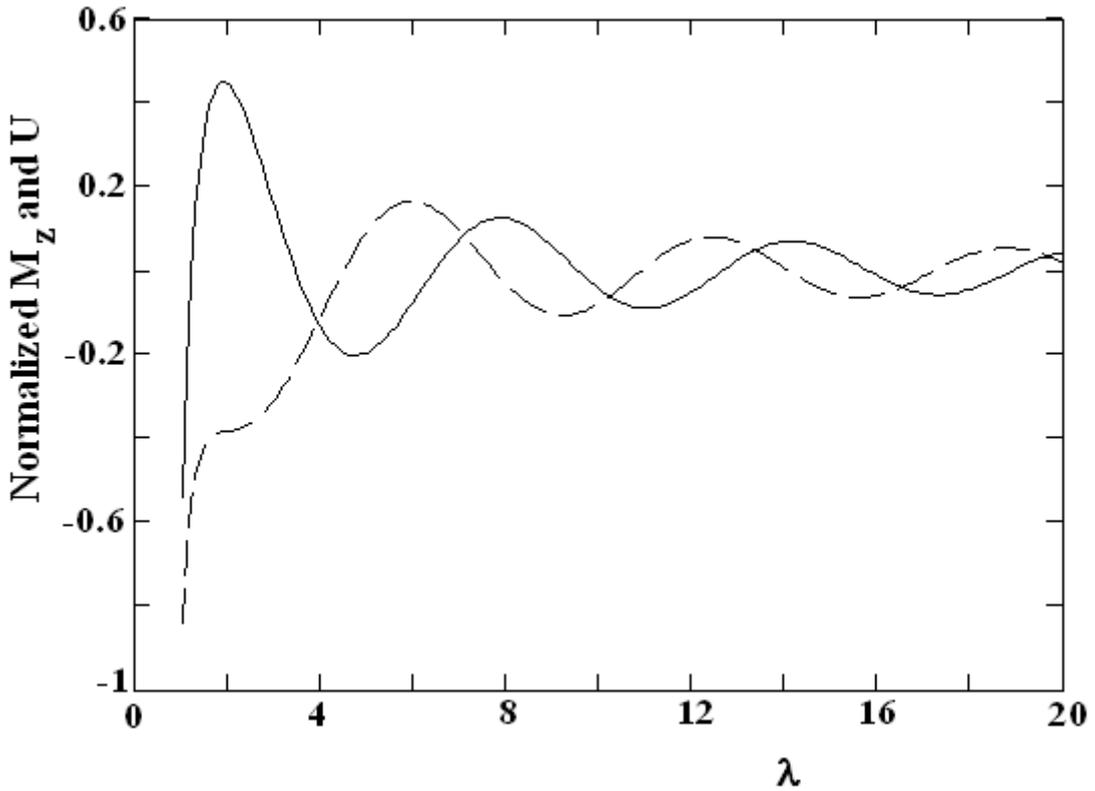

Fig. 2 Normalized values of $M_z$ (solid line) and $U(\lambda)$ (dashed line).

According to (21), (22), the dynamics of rotating dipole particle near the surface depends on the temporal evolution of the retardation parameter $\lambda = 2\Omega z_0/c$ in a complex manner, since the particle motion in the direction toward the surface and in the opposite direction affects the angular velocity and vice versa.

*Dielectric and metallic surface*

In order to examine the role of material factors, we also performed numerical calculations of $U(z_0)$ and $M_z$ according to Eqs. (12) and (15) in the case of dielectric ($SiO_2$) and metallic ($Au$) surface. The corresponding dielectric functions were taken from [11]. So, in the case of $SiO_2$ we used the dielectric permittivity

$$\varepsilon(\omega) = \varepsilon_\infty + \sum_{j=1}^{2} \frac{\sigma_j}{\omega_{0,j}^2 - \omega^2 - i\omega\gamma_j} \tag{23}$$

$\varepsilon_\infty = 2.0014, \sigma_1 = 0.00193(eV)^2, \omega_{0,1} = 0.057eV, \gamma_1 = 0.00217eV, \sigma_2 = 0.0102(eV)^2, \omega_{0,2} = 0.133eV$
$\gamma_1 = 0.00217eV, \sigma_2 = 0.0102(eV)^2, \omega_{0,2} = 0.133eV, \gamma_2 = 0.00551eV$.

For $Au$ we used the Drude model with $\omega_p = 1.37 \cdot 10^{16}\ rad/s$, $\tau = 1.89 \cdot 10^{-14}\ s$:

$$\varepsilon(\omega) = 1 - \frac{\omega_p^2}{\omega(\omega + i/\tau)} \tag{24}$$

Figures 3—5 show the results of our calculations for the reduced values of energy $U(z_0)$ and torque $M_z$ in dependence of $z_0$ and $\Omega$. Figure 3 corresponds to a $SiO_2$ slab, and Figs. 4,5 correspond to an $Au$ slab. The angle of inclination of the dipole was assumed to be $\theta = \pi/2$. The dotted lines on Figs.3a, 5a and the dashed line on Fig. 4a correspond to the nonretarded formula (12a). The torque in the nonretarded case equals zero.

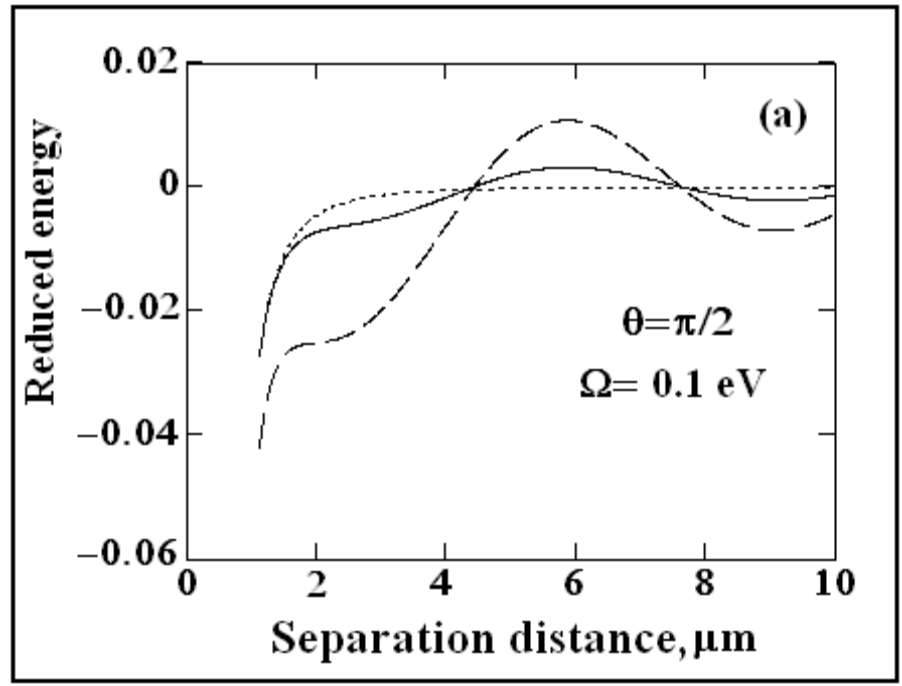

Fig. 3a Dependence of the reduced energy of rotating dipole on the separation distance near the surface of $SiO_2$. Solid, dashed and dotted lines correspond to Eqs. (12), (22) and (12a).

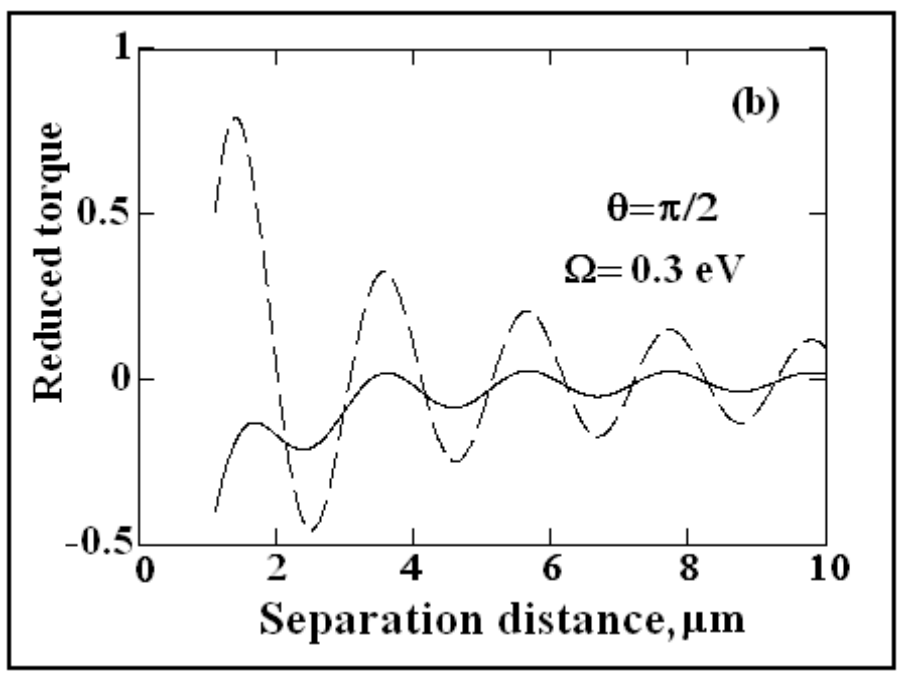

Fig. 3b Dependence of the reduced torque on a rotating dipole on the separation distance near the surface of $SiO_2$. Solid and dashed lines correspond to Eqs. (15) and (21).

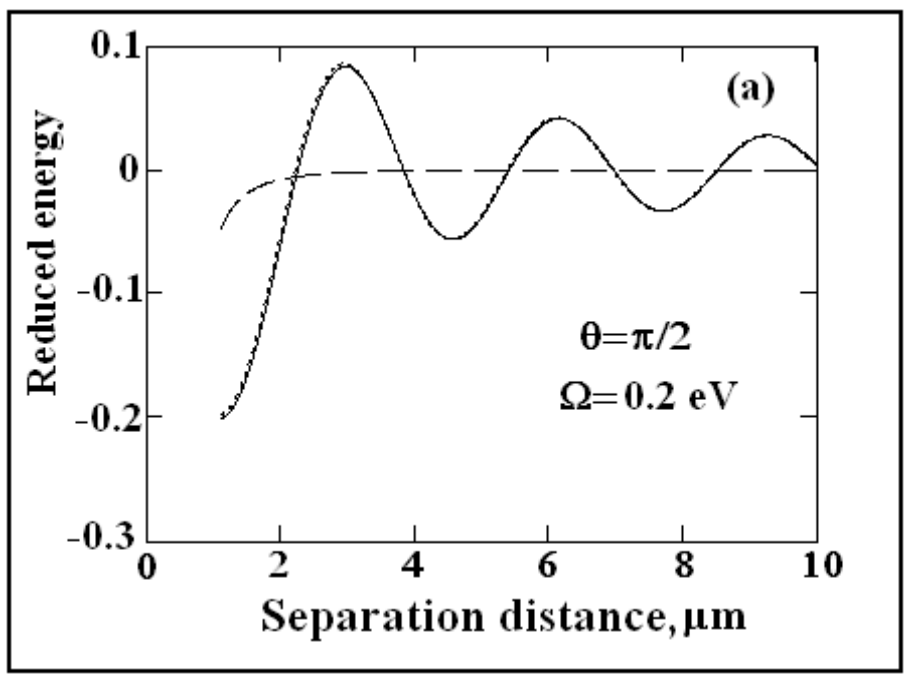

Fig. 4a same as on Fig. 3a in the case of $Au$ slab (low-frequency case)

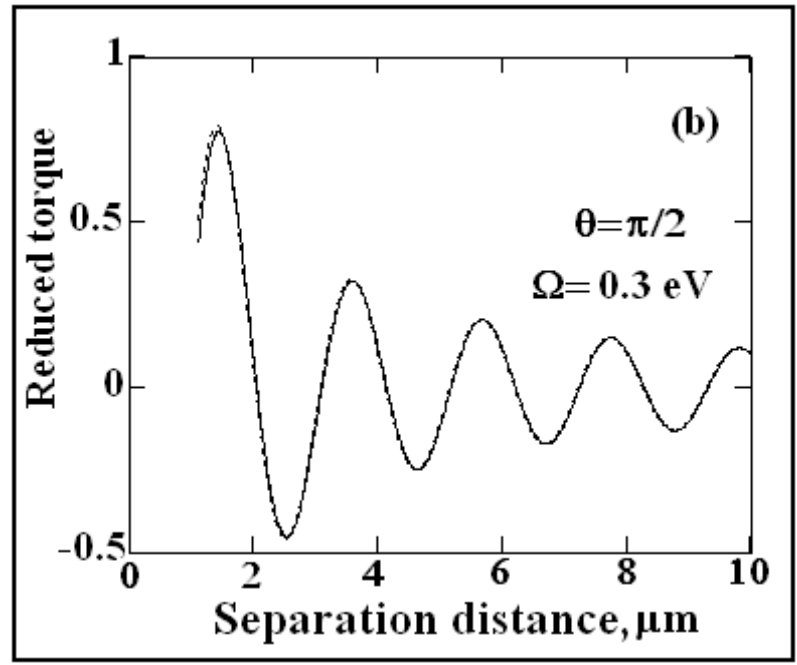

Fig. 4b Same as on Fig. 3b in the case of Au slab (low-frequency case).

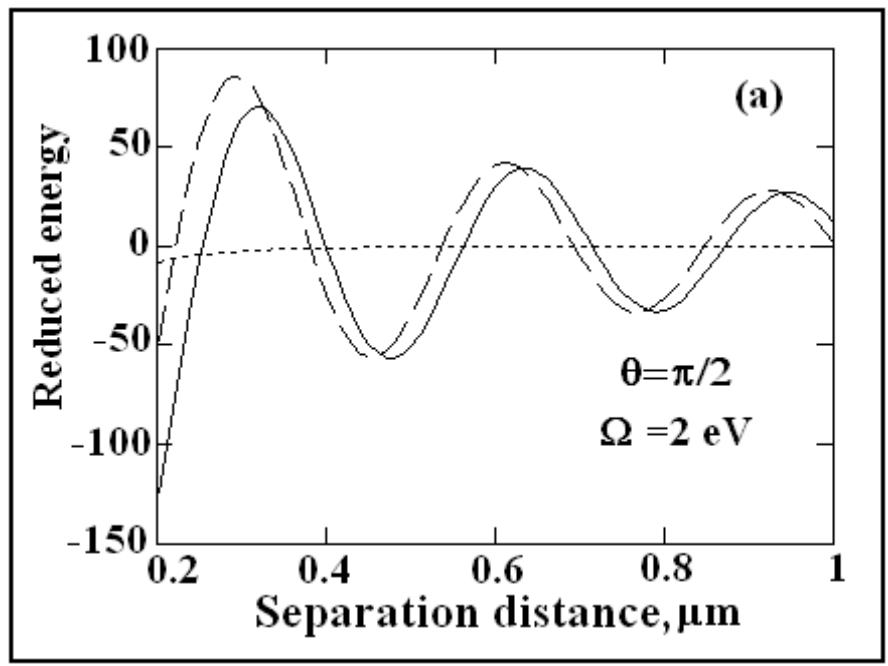

Fig. 5a Same as on Fig. 4a (high-frequency case).

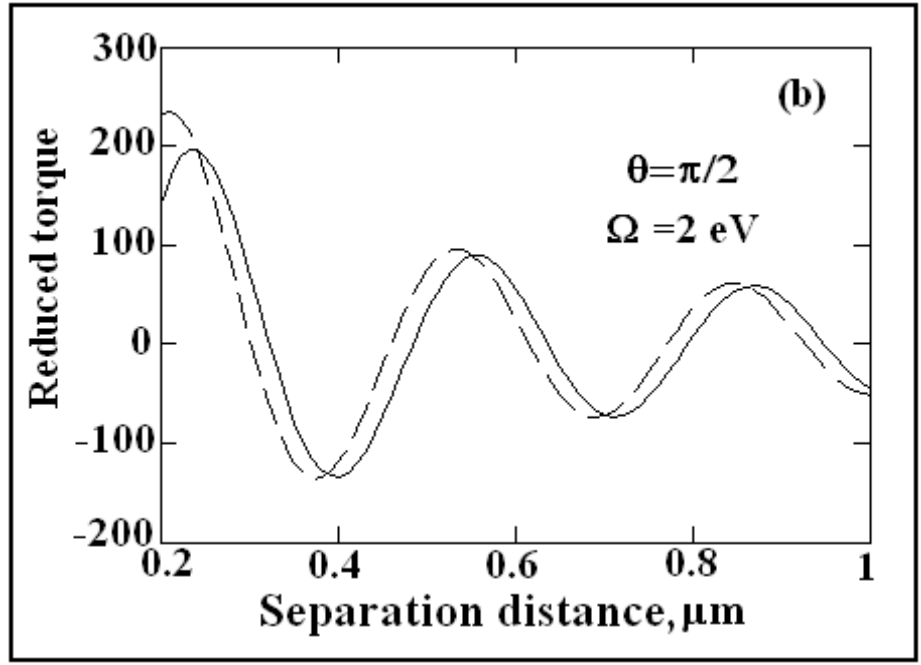

Fig. 5b Same as on Fig. 4b (high-frequency case).

Similar to Fig.2, from Figs. 3—5 it follows that the interaction energy and torque of rotating particle can be both positive (corresponding to repulsion and acceleration) and negative (corresponding to attraction and friction). In the case of metal slab at low frequencies, the curves shown by the solid lines (Drude metal) are very close to the dependences corresponding to the case of an ideally conducting surface (dashed lines). The deviations from an ideal surface become appreciable only at higher rotation frequencies (Figs. 5). At small values of $z_0$ and $\Omega$ (Fig.3a), the role of material properties becomes less significant, though in this case (due to a finite value of $\varepsilon(0)$), the difference between the dashed and dotted (solid) lines still will be observed.

**4. Conclusions**

We have examined the impact of rotation on the interaction of a small dipole particle with a surface of dielectric. We have calculated the energy of interaction and the torque acting on the particle. As it follows from the obtained expressions, rotation of the particle does not explicitly result in the time dependence of the potential energy and the $Z-$component of torque in the case where the rotation axis coincides with the surface normal. The torque involves a frictional component, that decelerates general rotation around the $Z$ axis, and an orientational component which tends to turn the dipole axis in the direction normal to the surface. If the dipole axis is directed exactly normal to the surface, the torque is zero and has no impact on rotation and orientation of the dipole, while the force of interaction with the surface coincides with the static (attractive) force. Therefore, the dipole may rotate for infinitely long time provided that the distance to the surface is the same. In the case where the dipole axis is inclined at an angle of $\pi/2$ relative to the surface normal, rotation of the particle has a noticeable effect on the interaction energy (force) and torque which may change the sign at some separations from the surface. So, the interaction of rotating dipole with the surface may result in a repulsive force and acceleration of the dipole rotation. We have estimated the corresponding effect in the case of an ideally conducting slab, and the $SiO_2$ ($Au$) slabs with the properties of dielectric and Drude metal.